\documentclass[aps,twocolumn]{revtex4-1}

\usepackage{graphicx}
\usepackage{wrapfig}
\usepackage{natbib}

\newcommand{\be}{\begin{equation}}
\newcommand{\ee}{\end{equation}}
\newcommand{\nn}{\mbox{} \nonumber \\ \mbox{} }
\newcommand{\ba}{\begin{eqnarray}}
\newcommand{\ea}{\end{eqnarray}}
\newcommand{\om}{\omega}

\newcommand\eg{{\it{e.g.\ }}}

\newcommand{\Bf}{{magnetic field}}
\newcommand{\Bfs}{{magnetic fields}}

\newcommand{\EM}{electromagnetic}

\begin{document}

\title{Electromagnetic ghosts   in pair plasmas}

\author{ Maxim Lyutikov \\
Department of Physics and Astronomy, Purdue University, \\
 525 Northwestern Avenue,
West Lafayette, IN
47907-2036 }

\begin{abstract}
Collisions of two  weakly  nonlinear, $a_0 \ll 1$, counter-propagating   EM pulses in pair plasma leave behind a  long-surviving  collection of localized   waves, 
{\it  an  electromagnetic ghost}.
 Waves are trapped (localized) by the  random large density fluctuations created by the beat between the  pulses. The process is  similar to random plasma density grating and/or   Anderson-like wave  localization. Structures survive for long, mesoscale times, while  the EM energy slowly bleeds through high density walls of the  density trap. Large  guide \Bf, $\om_B \geq $ few $\om$,  suppresses  the  formation of the ghosts.
 \end{abstract}

\maketitle

\section{Introduction}

The present work touches on a number of issues in the physics of  high intensity lasers, and plasma astrophysics.
Modern powerful lasers can accelerate particles to relativistic velocities, resulting in the production of $e^\pm$ pairs \citep{2006RvMP...78..591M,2008PhRvL.101t0403B,2012PhRvL.108p5006R,2020PhPl...27e0601Z}. The physics of ultra-strong laser-matter interaction also became a forefront research topic in relativistic plasma astrophysics, initiated by the meteoritic developments over the last years in the field of mysterious Fast Radio Bursts (FRBs) - ultra-intense millisecond-long radio bursts coming halfway across the visible Universe \citep{2007Sci...318..777L,2016MNRAS.462..941L,2022A&ARv..30....2P,2019ARA&A..57..417C,2016Natur.531..202S}.
Finally, nonlinear interaction of EM beams creates plasma  density  grating that is used in  laser pulse compressor  schemes \citep{2022PhRvP..18b4026E,2024PhRvE.110a5209L}. 

The present work follows on \citep{2025arXiv250906230T,2025arXiv250920594L} in investigating nonlinear effects in pair plasma. The key point is that in pair plasmas  the density  grating resulting from the nonlinear wave interactions produces charge-neutral density structures. Since there is no Coulomb repulsion, the resulting density fluctuations turn out to be much larger than in the electron-ion  plasmas. 

In this work we  first    demonstrate that interaction of weakly nonlinear waves can produced long-lived density structures in pair plasma - electromagnetic ghosts. Their dynamics occurs on mesoscales - intermediate between small kinetic and macroscopic scales. After investigating the properties of the  electromagnetic ghosts, we then argue that a pair  plasma subject to mildly intense \EM\ waves becomes ``granular'' - a highly density inhomogeneous medium.

\section{\EM\ ghosts in 1D}

\subsection{The code and parameters}

The simulations were performed using the EPOCH code \cite{Arber:2015hc}. Both   boundary condition are ``simple laser'' (for the EM fields coming from the inside of the plasma, this is equivalent to ``open'' boundary condition).

Our parameters are: both lasers' wavelengths $\lambda =10^{-4}$ cm; plasma density $n/n_{cr} = \om^2/\om_p^2=10^{-2}$ ($\om_p$ is defined with respect to each component separately);  slab thickness is $100 \lambda$. Laser intensity is parametrized by parameter $a_0$
\be
a_ 0  = \frac{e E_w}{m_e c  \om} =10^{-2}
\ee

The particular choice of $a_0$ is bounded by two factors: wave localization (formation of ghosts) increases with $a_0$, yet for larger $a_0$ , the pair  plasma is swept by the EM pulse (\eg, for  as little as  $a_0 =0.03$ the density increase is of the order of unity \citep{2025arXiv250920594L}. 

The strongest effect is when both pulse are circularly polarized  (CP), and in  the PLUS configuration, with the same sense of rotation in absolute space (not with respect to there wave vector, see  \citep{2025arXiv250920594L}.)  In the  opposite case, MINUS configuration, there is no ghosting.

A nonlinear pulse is expected to be  self-localized on scales  \citep{2025arXiv250920594L}
\ba &&
L\approx    \frac{1}{\rho_L} \frac{c}{\om}
\nn &&
\rho_L =   \left(    a_0 \frac{  \om_{p} }{\om }\right)^{2/3}  \approx 100,
\label{self} 
\ea
(numerical estimate is for the parameters of these simulations), 
see Appendix \ref{self-localization}.
Since we are interested in the effects arising due to interaction of the beams, we limit the pulse width    to $ \leq L$.

For 1D runs, the code parameters are set to $n_x =100$ (number of cells per wavelength) and $n_p =100$ (number of particles per cell). Unless specified,
the initial configuration is cold plasma, see \S \ref{T} for discussion of thermal effects. 

\subsection{The \EM\ ghost: basic  1D results (cold plasma, no \Bf)}

We first discuss in details the properties of the 1D simulations, and  later, in \S \ref{2d} we explore the 2D beams, modulated in transverse and longitudinal directions

\begin{figure}[h!]
   \includegraphics[width=.99\linewidth]{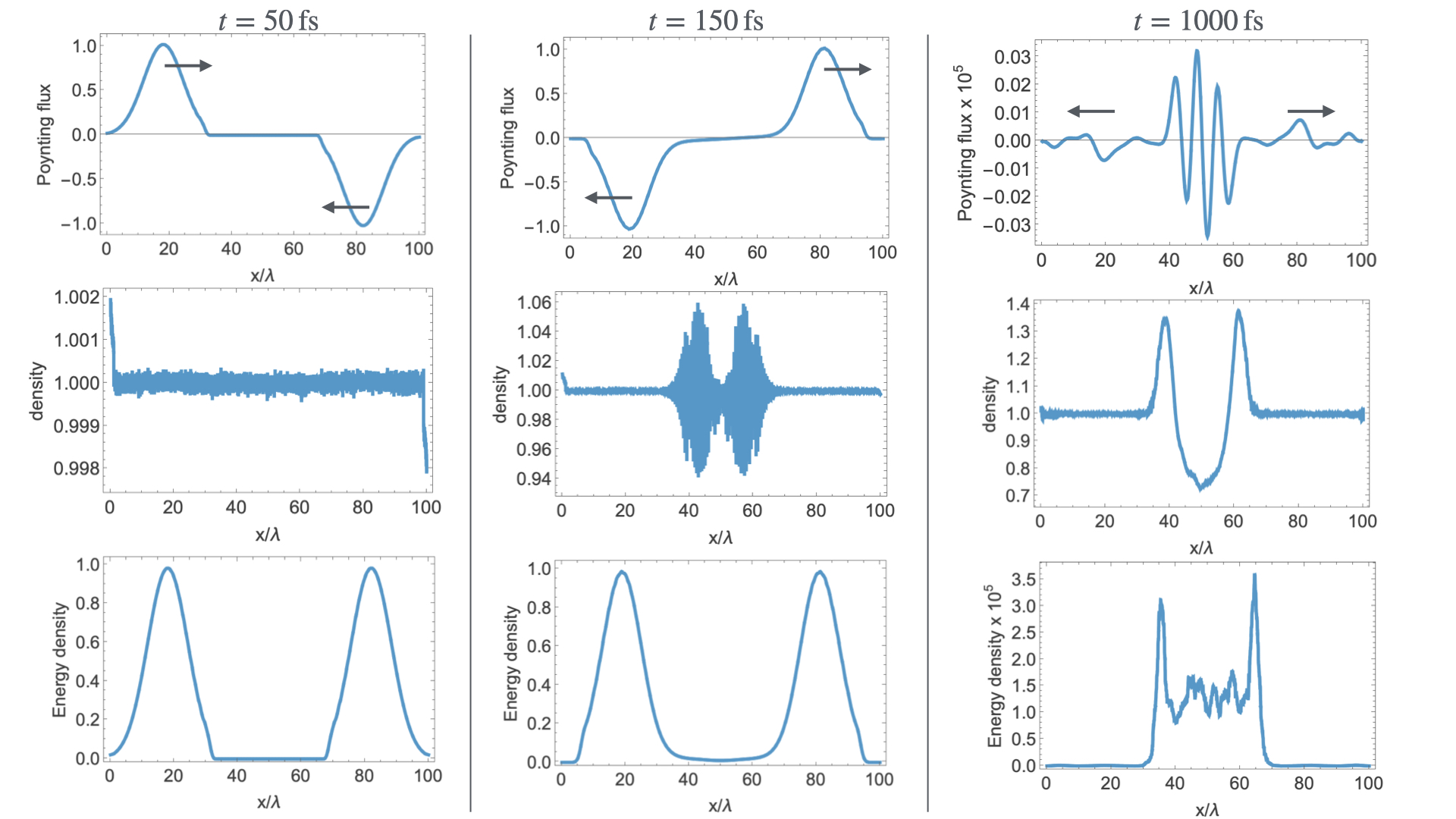}
\caption{Colliding EM pulses in pair plasma. Columns (left to right): snapshots at time  $50$ fs (before collision), $150$ fs (soon after collision), $1000$ fs (long after collision). Rows (top to bottom): Poynting flux, density, energy density. 
 Poynting flux and energy density are normalized to peak values, except in the right column where they are multiplied by $10^5$.
 All plotted quantities are averaged over one wavelength. Peak nonlinearity $a_0=10^{-2}$, pulses' duration full width at half max is 50 fs. 
  $n_x=100,\, n_p=100$. 
 (Results for $n_x=30,\,  n_p=30$ look qualitatively similar.) }
\label{main}
\end{figure}

Our main result is featured in Fig. \ref{main}. We plot snapshots of Poynting flux (top row, normalized to peak value of the pulses), plasma density  (middle row, normalized to the undisturbed value) and plasma \EM\ energy density (bottom row, also normalized to peak value of the pulses). Two CP with PLUS relative polarization.

Before the collision (top panel), the CP  pulses do not disturb plasma much. During the interaction, the beat  between the laser pulses produces large, random density fluctuations  that survive for times much longer than the interaction time (central panel). Resulting random  density  fluctuations create random plasma density grating, and corresponding random fluctuations of the dielectric prematurity  $\epsilon$.  These random fluctuations of $\epsilon$ Anderson-localize the \EM\ energy (bottom panel), creating the \EM\ ghost. The ghost survives for times much longer than the interaction time, and actually, it survives for mesoscopic time scales, much longer since the \EM\ pulses left the simulation domain (bottom panel). 

The \EM\ energy is ``sloshing'' within the ghost, eventually creating well-structured density wall. The amplitude of density perturbations first even increases with time (compare middle and right panel in the middle row). The energy is slowly leaking from the ghost, as indicated by two arrows in the top right panel (one can check that the average value of Poynting flux is negative in the left half of the simulation box and positive in the right half). Drainage is not continuous, but time dependent/oscillatory, as indicated by wave-like structures of the Poynting  flux outside the ghost. 
While at late times the density and  energy density evolve slowly, the Poynting flux shows  sloshing waves within the ghost, penetrating the density barrier,   and propagating outwards.

The size of the ghost is approximately two times the size of each pulse -  this is the interaction length of the pulses. 

Since the energy density in the ghost is $\sim 10^{-4}$ of the peak energy density of the main pulses, relatively long simulations are needed to see the ghost,  the main pulses need leave the box way before the ghost shows up.

\begin{figure}[h!]
 \includegraphics[width=.8\linewidth]{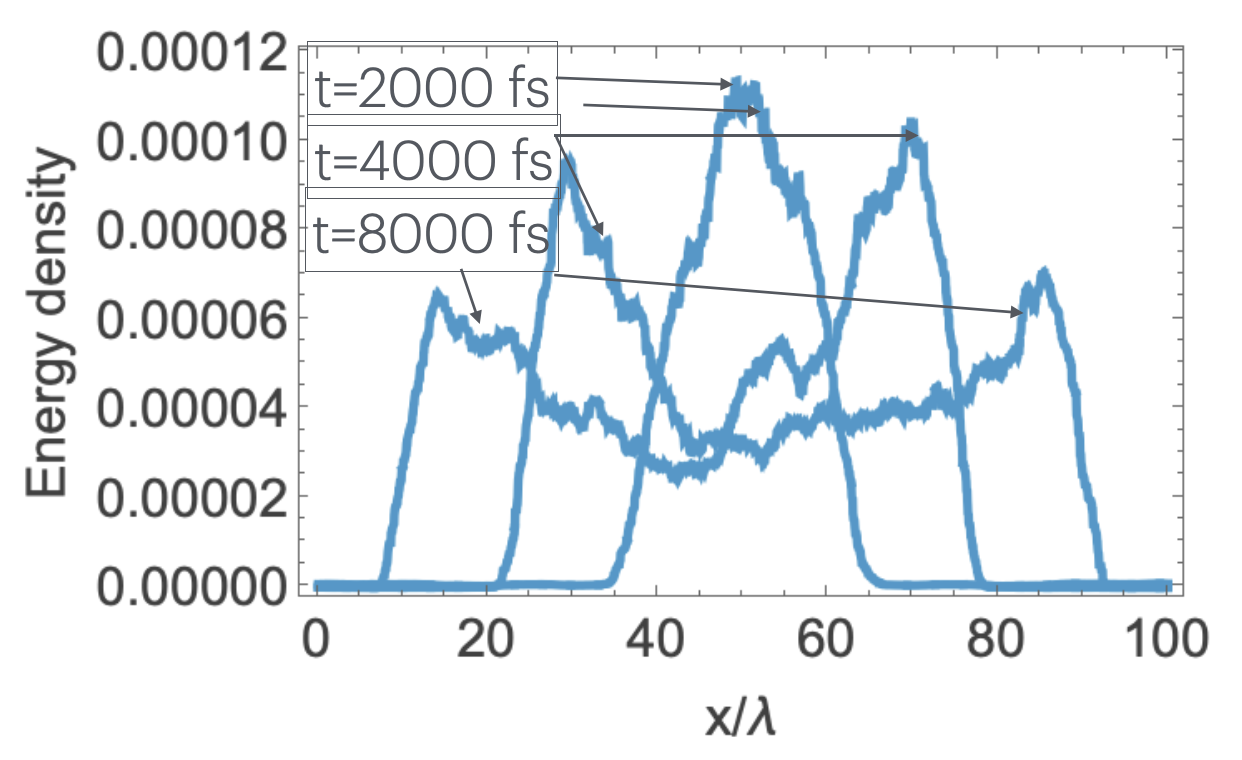}
  \includegraphics[width=.8\linewidth]{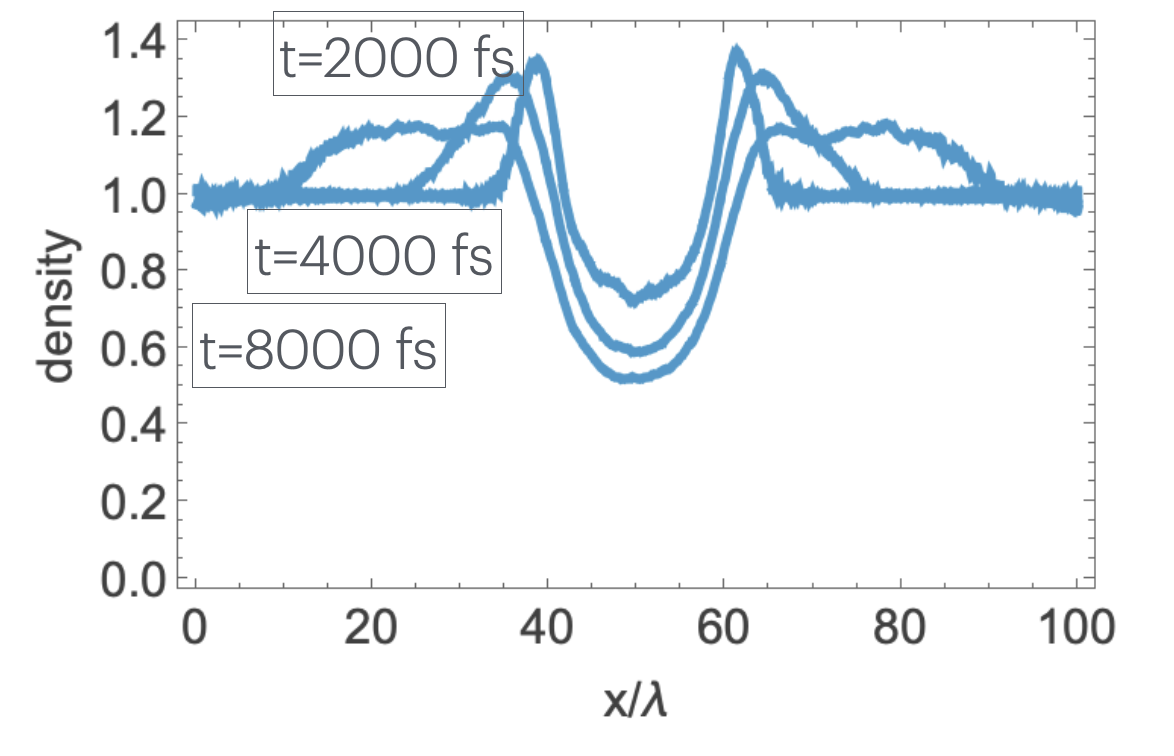}
\caption{Evolution of the structure of the ghost with time:  profiles of energy density (top panel) and plasma  density (bottom panel). With time the ghost becomes wider. With time  the central density depletion  increases.  In these simulations $n_x=n_p=100$. 
}
\label{1d-time}
\end{figure}

In Fig. \ref{1d-time} we show long-term evolution of the ghost. The ghost slowly expands, remaining as a coherent structure for a very long time. 

\subsection{Guide \Bf,  plasma temperature, polarization effects}

\subsubsection{Effects of guide \Bf}
It is convenient to introduce parameter
\be
b_0 = \frac{\om_B}{\om} 
\ee
$b_0=1$ corresponds to cyclotron resonance (and efficient absorption of the wave). 

Outside of the regime $b_0 \approx 1$,  the effects of the guide \Bf\ are as follows 
For $b_0 \leq 1$ (sub-resonant pulse), the strength of the ghost do not depend much on the values of the guide field  $b_0$. 
For $b_0 \geq 1$, the strength of the ghost  sharply decreases for  $b_0 \geq 2$. For $b_0 = 4 $ the ghost nearly disappears.

 We also comment that generally  one cannot  compare the properties of the ghost as a function of $b_0$ at fixed simulation time since the EM propagation speeds, interaction time and length, all  depend on \Bf. 
 
For $b_0 \geq 1$, the strength of the ghost decreases with $b_0$.

 \begin{figure}[h!]
  \includegraphics[width=.8\linewidth]{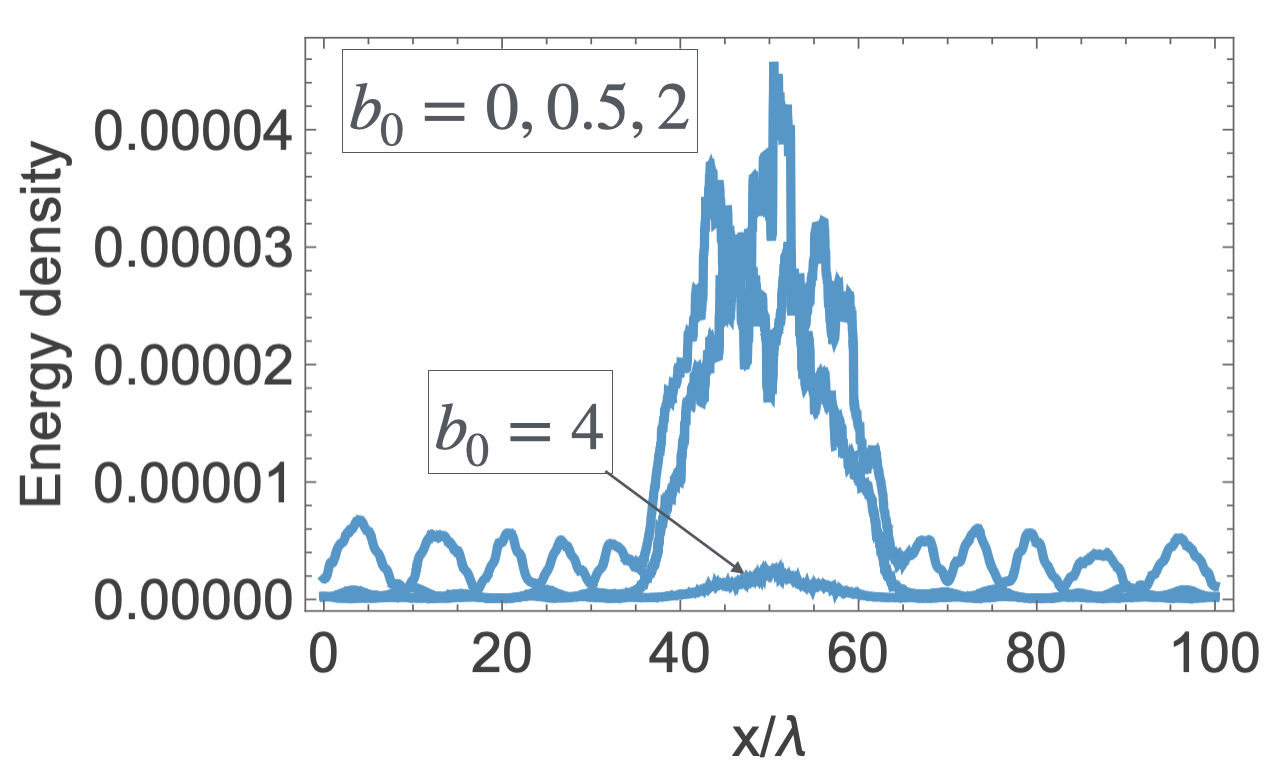}
   \includegraphics[width=.8\linewidth]{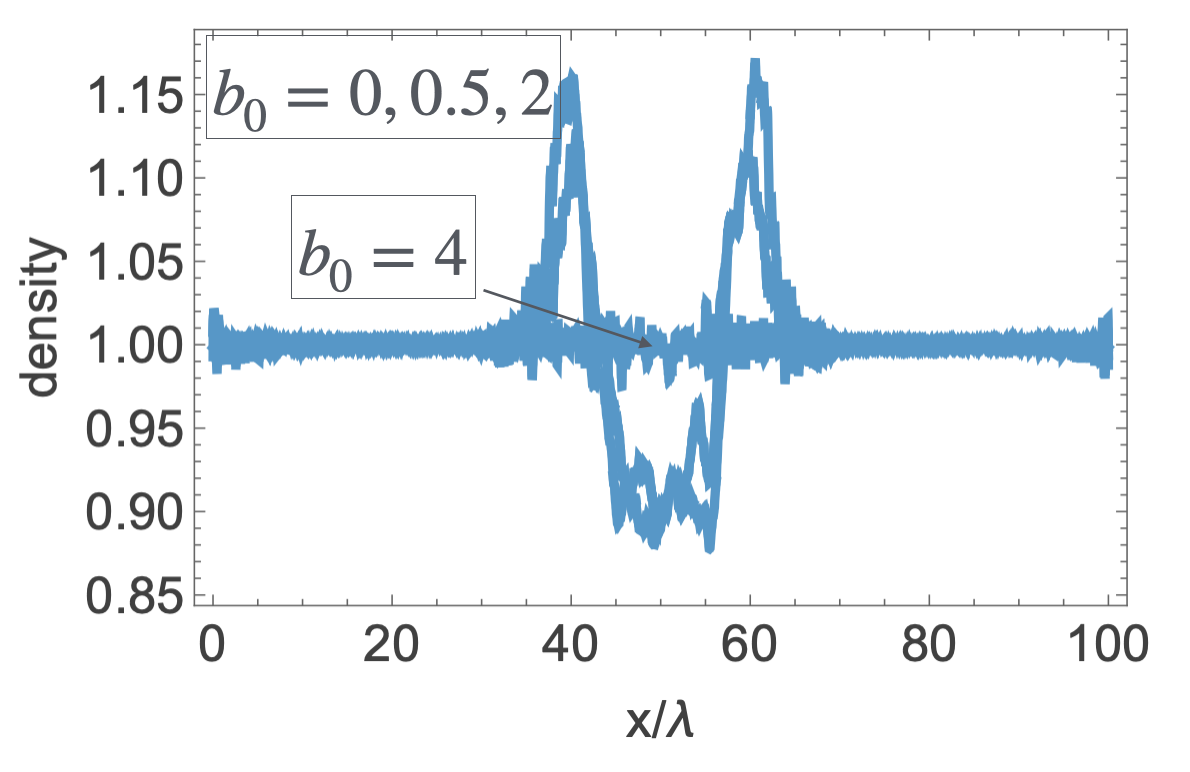}
\caption{Dependance of the ghost on guide \Bf. For $b_0=0, \,  0.5, \,2$ the curves are almost coincident. For $b_0=4$ the ghost nearly disappears. In these simulations $n_x=n_p=100$. 
}
\label{b0}
\end{figure}

\subsubsection{Initial temperature}
\label{T}
\begin{figure}[h!]
 \includegraphics[width=.8\linewidth]{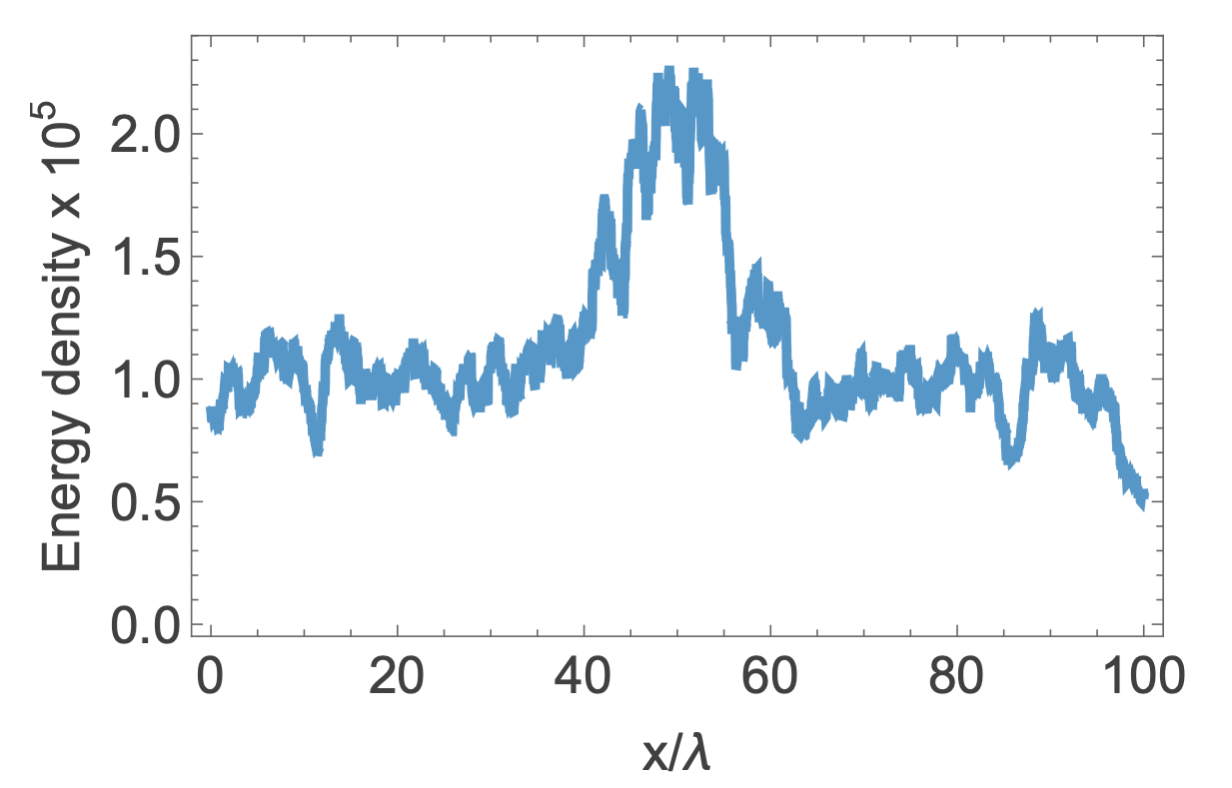}
  \includegraphics[width=.8\linewidth]{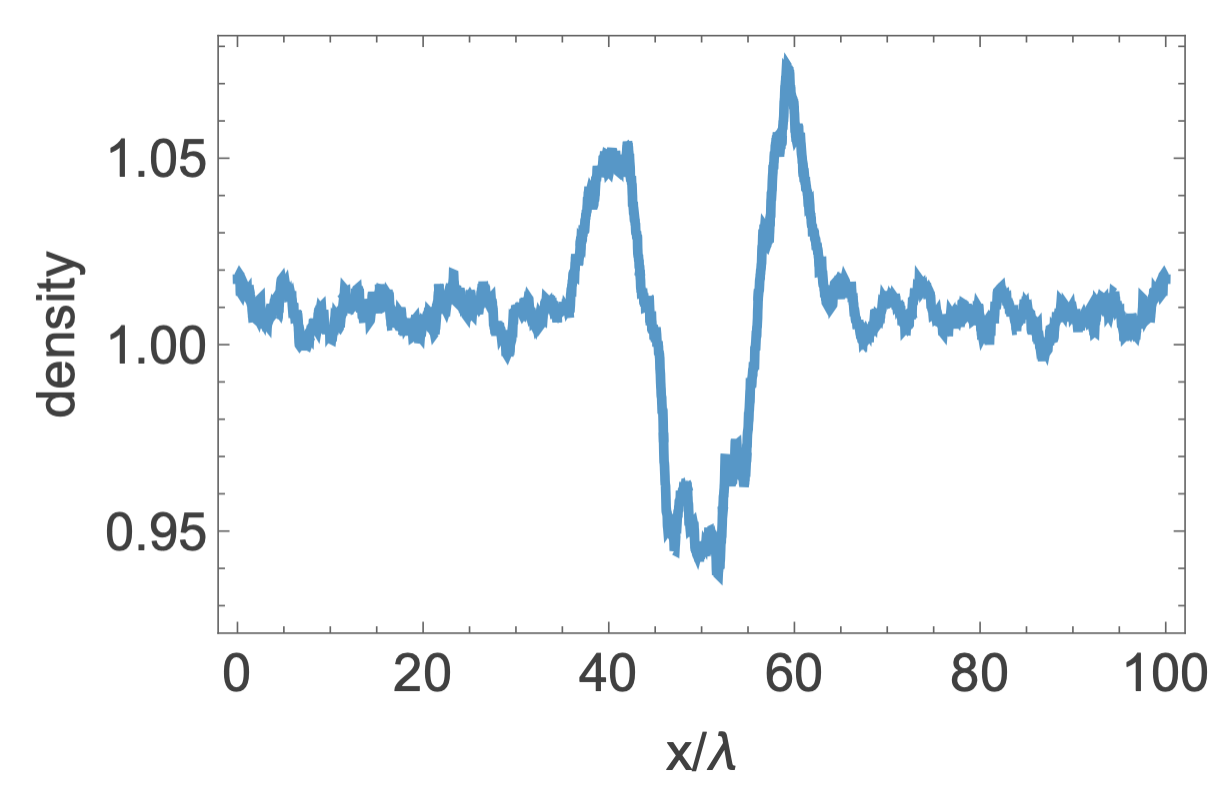}
\caption{Effect of temperature. At $\Theta = 0.5 \times a_0^2 = 5\times 10^{-5}$ the ghost is nearly gone (and even weaker for larger temperatures). 
}
\label{Theta}
\end{figure}

For initially hot plasma, the  physically  relevant parameter is 
\be
\Theta_0\equiv \frac{k_B T}{m_e c^2} = a_0^2
\ee
For  $\Theta> \Theta_0$ the thermal velocity of particles exceeds the jitter velocity in the wave. This leads to decoherence, and ghosts disappear. Our numerical results are in agreement with this prediction. In  Fig. \ref{Theta} we plot profiles of energy density for one particular value of  $\Theta = 0.5 \times a_0^2 = 5\times 10^{-5}$. The ghost is nearly gone.

 Investigation of a particular dependence on temperature  (at smaller $\Theta$) requires  heavy calculations. Due to numerical heating, initially cold plasma achieves condition $\lambda/n_x \approx 10  r_D$  \citep[$r_D$ is Debye radius,][]{birdsall}. The corresponding numerical temperature is 
 \be
 \Theta_{num} \approx   \frac{1}{n_x^2} \left( \frac{n}{n_{cr}} \right)
 \ee
 Condition $\Theta_{num}  \leq \Theta_0$ then requires  resolution:
\be 
n_x \geq  \frac{1}{a_0} \left( \frac{n}{n_{cr}} \right)^{1/2} \approx 10
\ee
Our numerical results conform with this estimate.

\subsubsection{Density granulation}

\begin{figure}[h!]
  \includegraphics[width=.9\linewidth]{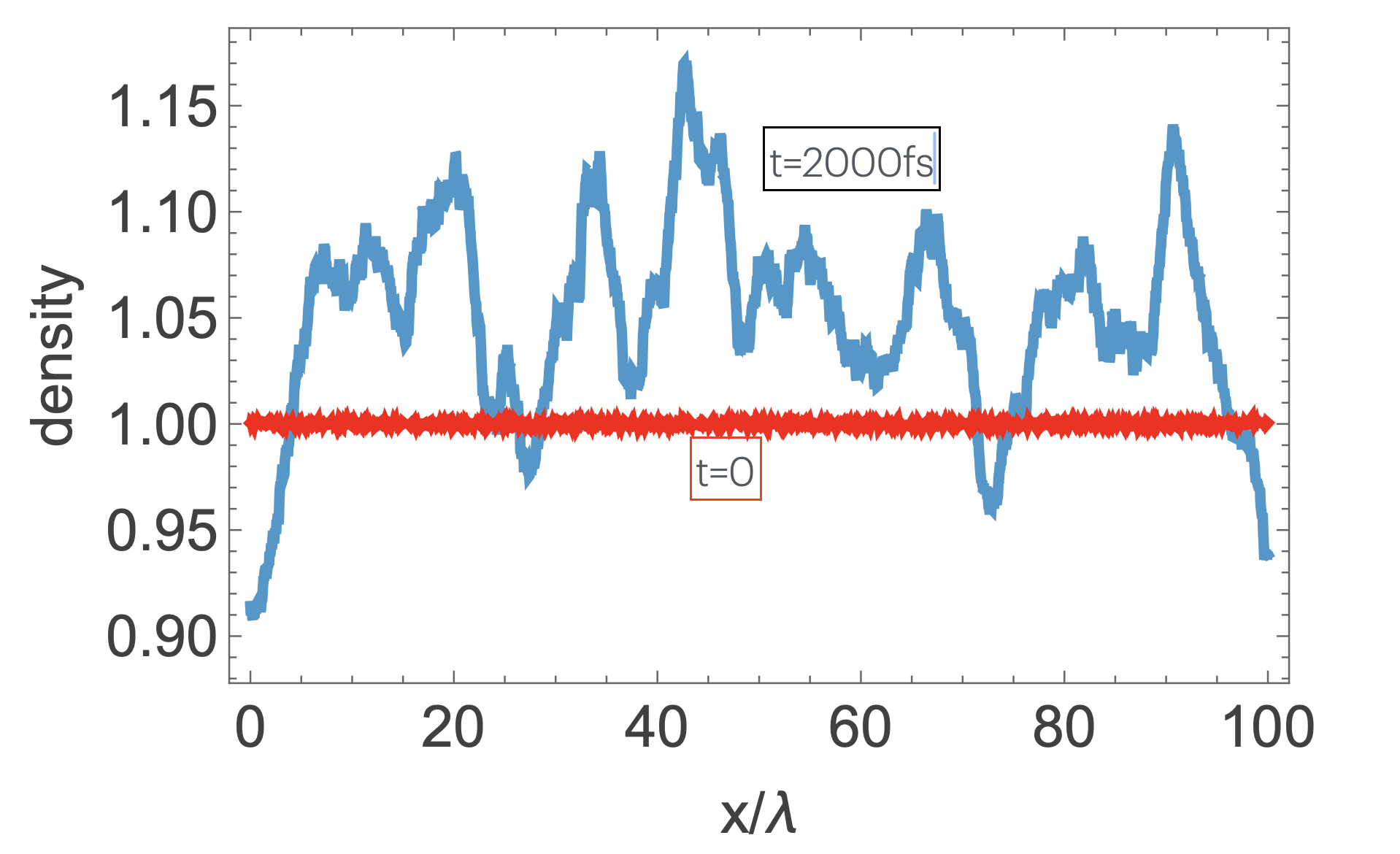}
  \caption{Density structure after collision of 5+5 pulses, at time $t=4000$ fs, compared with the undisturbed plasma. Density is  averaged over one wavelength. 
Slight overall increase is due to ponderomotive push   from the pulses.}
\label{granulation}
\end{figure}

Density perturbations induced by weakly nonlinear \EM\  waves survive in pair plasma for a very long time. As a result, pair  plasma subjected to a burst of \EM\ waves becomes ``granular'', with large density fluctuations,  even long after the pulses  exited the plasma. In Fig. \ref{granulation} we show density structure after  5 pulses from each side passed through (10 pulses total). Each pulse has a duration  of 50 fs, pulses are separated by 100 fs. The last pulse left the simulation domain around 500 fs - snapshot is  taken at 4000 fs.

\section{\EM\ ghosts in 2D}

\begin{figure}[h!]
  \includegraphics[width=.8\linewidth]{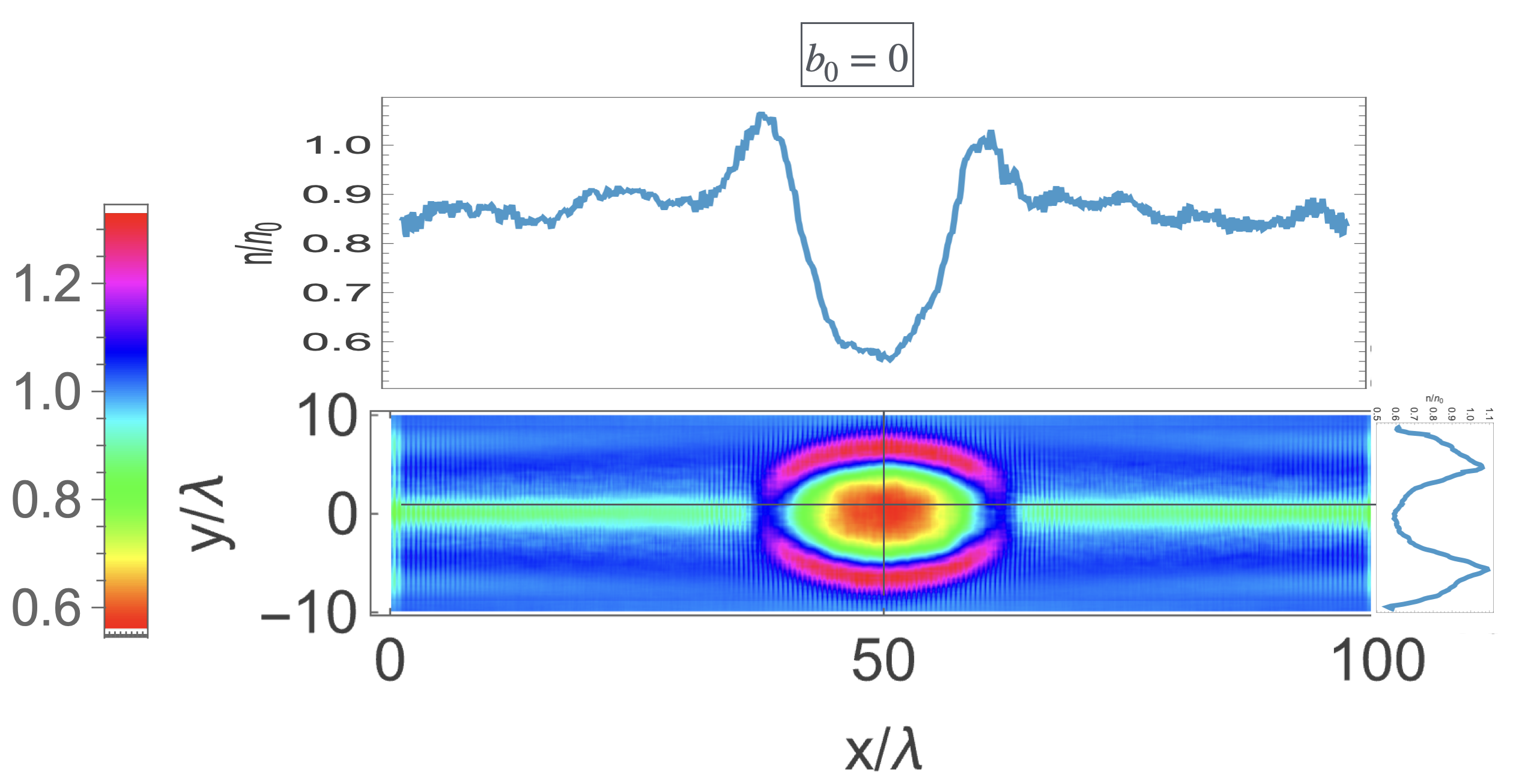}
  \includegraphics[width=.8\linewidth]{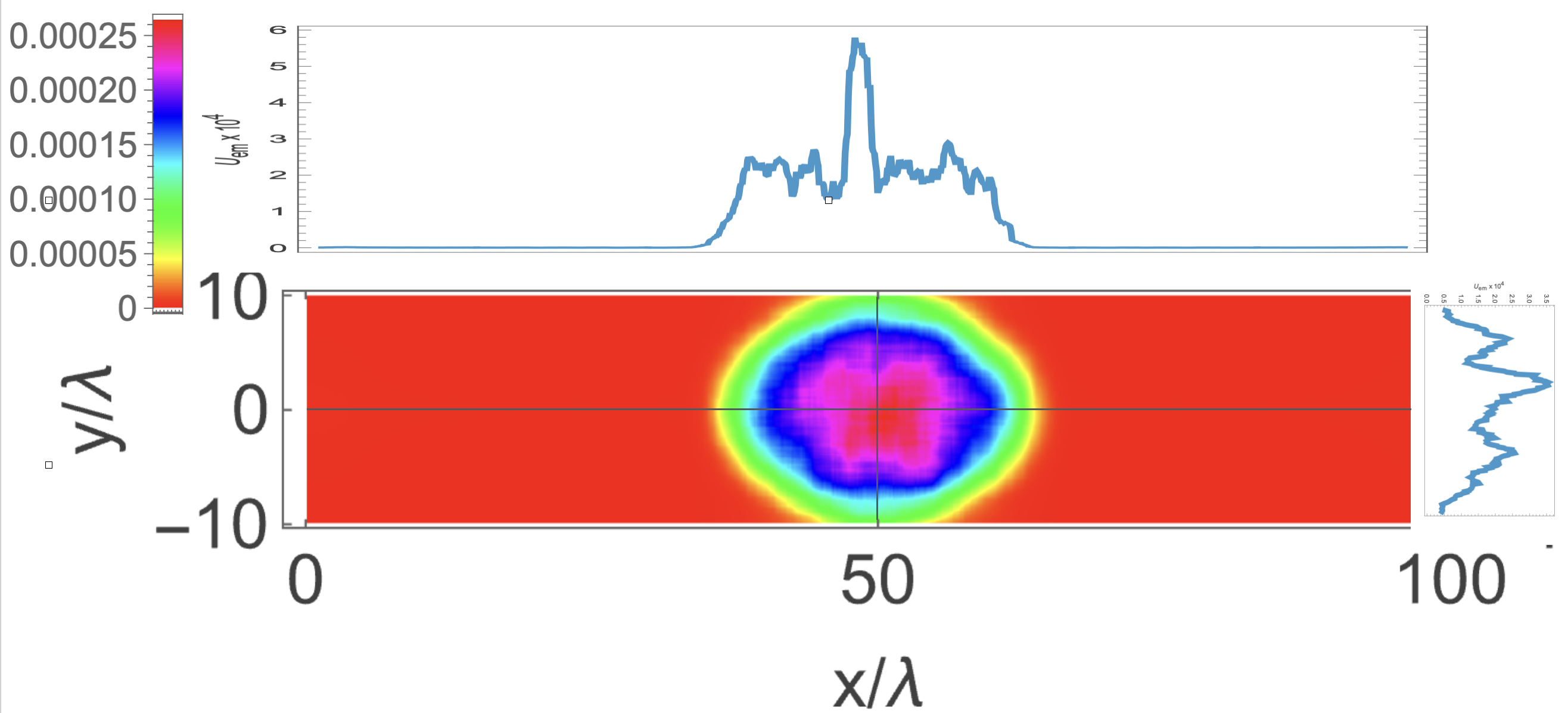}
  \caption{  2D simulations of \EM\ ghost, zero guide fields. Density  (top)  and \EM\ energy density (bottom), time 1000 fs; attached panels are middle cuts. Energy density is normalized to the peak energy density of the pulses; typical value within the ghost is $10^{-4}$ ($n_x= n_p=30$). 
}
\label{2duemb0}
\end{figure}

\begin{figure}[h!]

 \includegraphics[width=.8\linewidth]{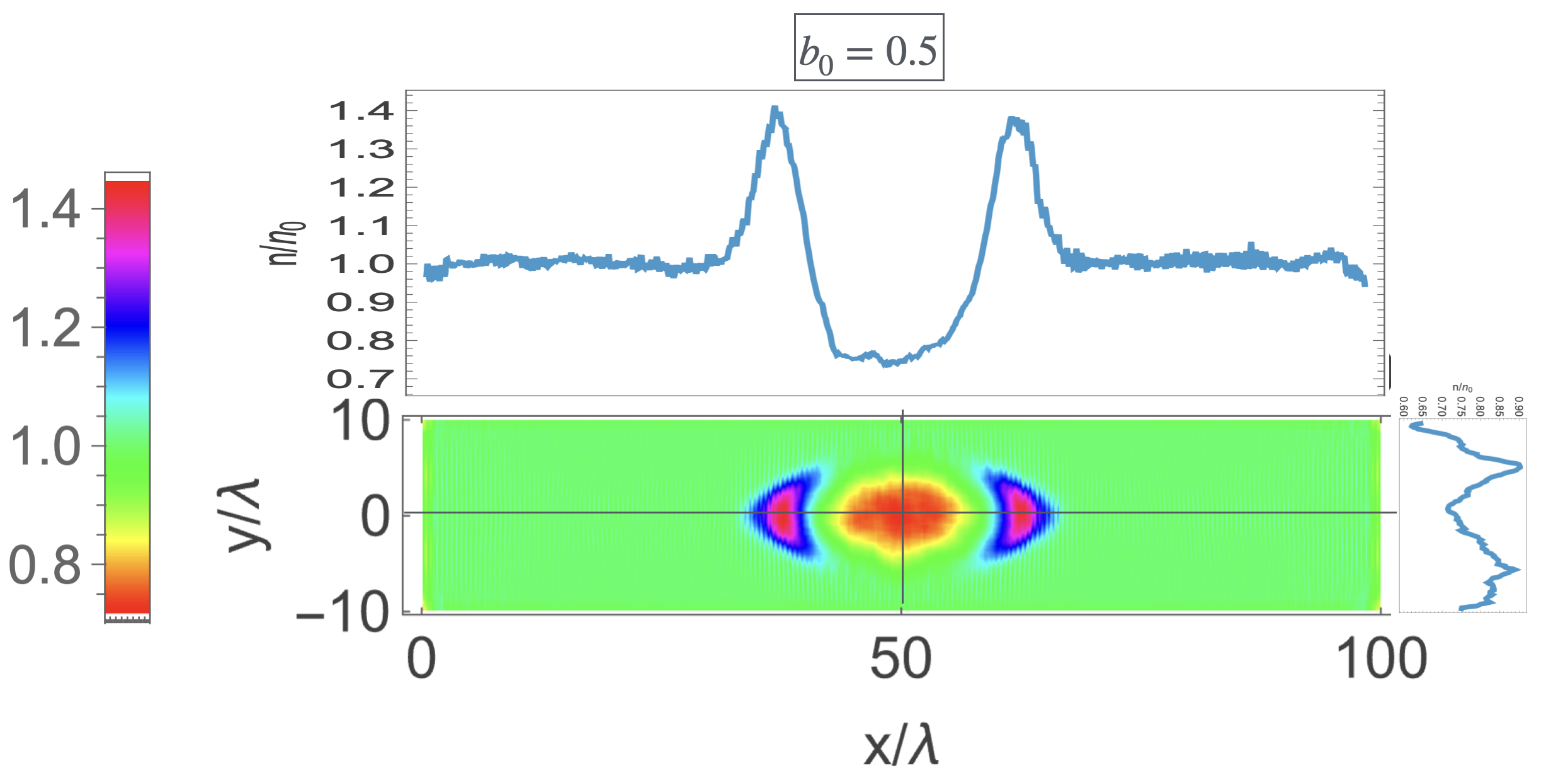}\\
  \includegraphics[width=.8\linewidth]{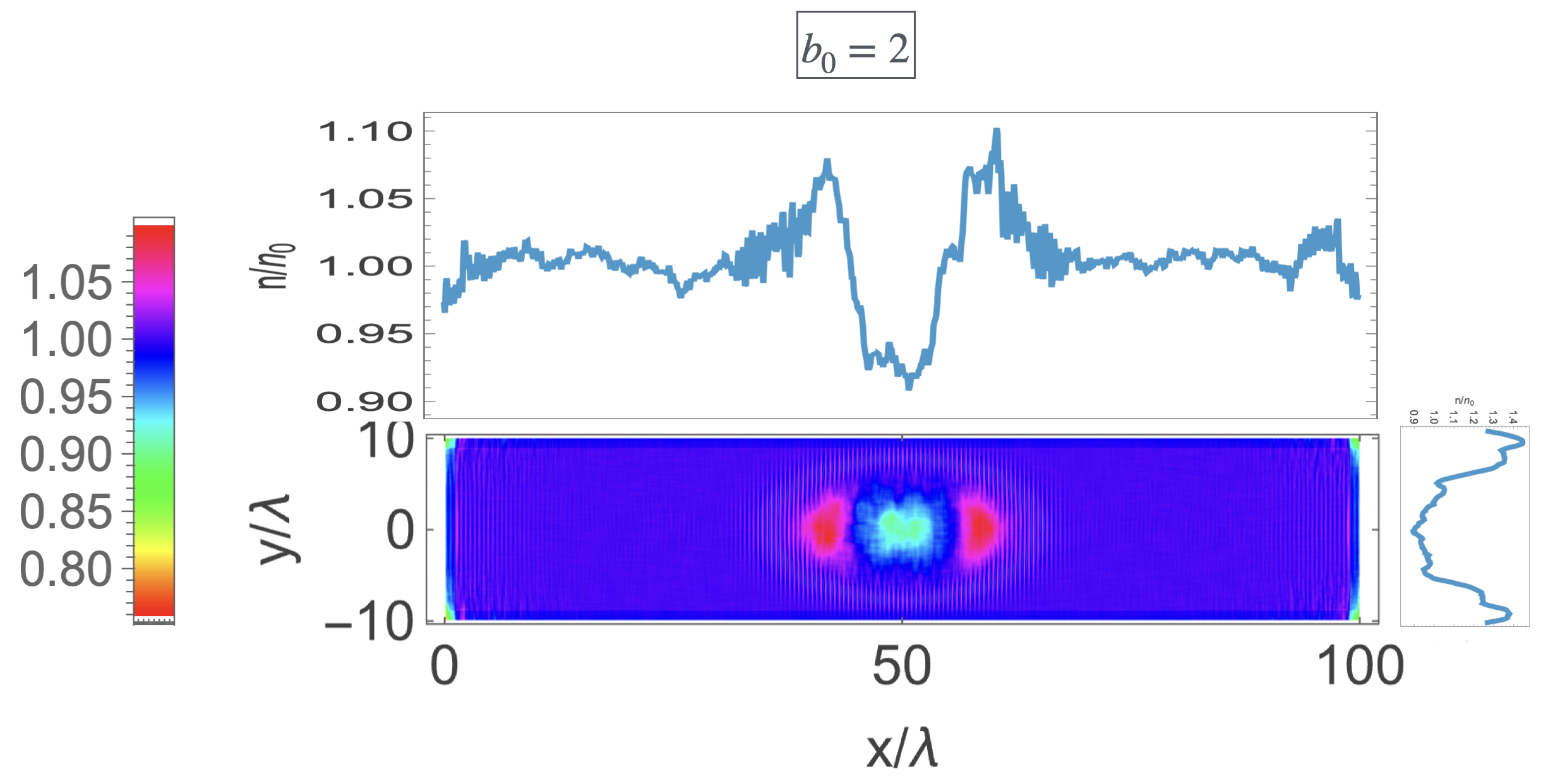}
\caption{2D simulations with guide \Bf\ field.  Depicted are plasma  density structures for  $b_0= 0.5,\, 2$, Top and right panels are middle cuts.}
\label{2d}
\end{figure}

We have run a number of 2D simulations. The transverse  Gaussian profile was set  to  $5 \lambda$, Figs. \ref{2duemb0}- \ref{2d}. The 2D simulations are generally consistent with the  1D simulations: an \EM\ cavity is formed. These 2D simulations importantly  demonstrate that the 1D  simulations/results  are  generic, not  specifically limited  to low  dimensionality. 

In Fig. \ref{2duemb0} we depict the  basic case, the  formation of the \EM\ ghost   following  an  encounter of two \EM\ pulses in unmagnetized plasma.  A  low density/high \EM\ density  2D cavity is formed.  Both the  transverse  and longitudinal extents match the expectation.


 Magnetized 2D cases, Fig. \ref{2d}, also follow the 1D simulations: cavities are formed for $b_0 \leq $ few. For sufficiently high guide \Bf\ ghosts also disappear in 2D.

\section{Discussion}

We discovered a new nonlinear plasma effect: \EM\ ghosts crated by colliding pulses in pair plasma. The effect we are after is different from  electrostatic echos/van Kampen waves \citep{Kadomtsev1968,1955Phy....21..949V}. In our case, the trapped waves are \EM.

There are several stages to trapping. At first, density fluctuations are random, appearing on scales of wavelength, and extend over the interaction length, approximately two times the size of the wave-packet.  This is the regime of Anderson localization \citep{1958PhRv..109.1492A}, or, more correctly, localization of light in random media
\cite{doi:10.1137/1104038,1991PhT....44e..32J}.  One may classify this as a linear trapping regime.  

As the original wave-packets leave, the \EM\ energy trapped inside the ghosts effectively pushes on the confining structures, sweeping plasma and leading to the generation of homogeneous density walls. One may classify this as a nonlinear trapping regime.  
With time,  the trapped \EM\ fields push the confining walls outwards. This, on the one hand, decreases the internal energy density of radiation, and, curiously, deepens the density well, bottom panel in Fig. \ref{1d-time}.

These effects persist for not too strong \Bfs, $\om_B \leq $ few $\om$.

We have also considered  linear polarization (LP).  For aligned case, when the polarization planes  coincide,  the ghost is also present, similar to the CP case.  Each LP pulse can be approximated by  two CP pulses, forming two pairs  of PLUS configurations. For orthogonal LPs there is no ghost.



This research was supported in part by grant NSF PHY-2309135 to the Kavli Institute for Theoretical Physics (KITP). 
I would like to thank  Paulo Alves,  Goetz Lehmann, Alexey Mohov, Alexander Philippov, Anatoly Spitkovsky and
Kavin Tangtartharakul.

\bibliographystyle{apsrev}

 \bibliography{/Users/lyutikov/Library/CloudStorage/Dropbox/Research/BibTex,/Users/lyutikov/Library/CloudStorage/Dropbox/Research/BibTexShort.bib,//Users/lyutikov/Library/CloudStorage/Dropbox/Research/NASA_FRB.bib} 
 
 \appendix

 \section{Anderson self-localization of long pulses}
\label{self-localization}

\begin{figure}[h!]
   \includegraphics[width=.89\linewidth]{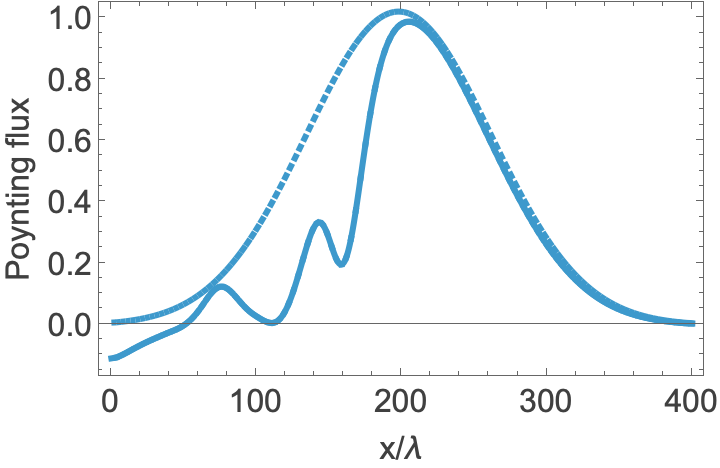}
\caption{Self-localization  of the pump. Plotted are Poynting fluxes, normalized to peak value, pulse duration 1000 fs. Dashed line is  vacuum profiles, solid line: pair plasma. The pump pulse in plasma is cut-off, approximately after  $\rho_L$, Eq. (\ref{self}). This is due to Anderson self-localization of the pump \citep{2025arXiv250920594L}. Notice negative values of Poynting fluxes at small $x$, indicating that energy is been drained from the pulse.
}
\label{self-loc-long}
\end{figure}

In pair plasma, long EM pulses experience Anderson self-localization. One of the consequences is that a sufficiently long pulse, exceeding (\ref{self}) is reflected from plasma \citep{2025arXiv250920594L}, Fig. \ref{self-loc-long}. 

This has  implications for schemes based on 
laser amplification  via stimulated Brillouin
scattering \citep{Cheriyan2022ComprehensiveReview,2013arXiv1311.2034A}, that  use two counter-propagating laser pulses. The length of the pump is usually a free parameter. In pair plasma it is not: it is limited by Anderson self-localization of the pump pulse \cite{2025arXiv250920594L}.  After traversing the length (\ref{self}) the pump self-localizes/reflects: in pair plasma the effective  pump duration is limited by (\ref{self}).   This limitation on Stimulated Brillouin
back-scattering  is specific to pair plasma. 

\end{document}